\documentclass[
    ,final            
  ]
  {aipproc}

\layoutstyle{8x11single}

\def\slash#1{\setbox0=\hbox{$#1$}  
   \dimen0=\wd0     
   \setbox1=\hbox{/} \dimen1=\wd1  
   \ifdim\dimen0>\dimen1   
      \rlap{\hbox to \dimen0{\hfil/\hfil}} 
      #1     
   \else     
      \rlap{\hbox to \dimen1{\hfil$#1$\hfil}} 
      /      
   \fi}      %

\newcommand{\Tr}{\mathrm{Tr}}

\newcommand{\bea}{\begin{eqnarray}}
\newcommand{\eea}{\end{eqnarray}}
\newcommand{\be}{\begin{equation}}
\newcommand{\ee}{\end{equation}}
\newcommand{\eg}{{\it e.g.}}

\begin{document}

       \title{Final state interactions \& the Sivers function}


\author{Leonard Gamberg}{
  address={Division of Science, Penn State University-Berks,
Reading, Pennsylvania 19083, USA}
}

\author{Marc Schlegel}{
  address={Institute for Theoretical Physics, Universit\"at  T\"ubingen, D-72076 T\"ubingen, Germany}
}

\begin{abstract}
The non-vanishing of naive T-odd parton distributions 
function can be explained   by
the existence of the gauge link which emerges from 
the factorized description of the deep inelastic scattering cross section
into perturbatively calculable  and non-perturbative factors.
This path ordered exponential describes initial / final-state interactions
of the active parton due to soft gluon exchanges with the target remnants.
Although these interactions are non-perturbative, studies of final state
interactions have been approximated by
 perturbative one-gluon exchange in Abelian models.
We include higher-order gluonic contributions from the
gauge link by applying non-perturbative eikonal methods, 
incorporating color degrees of freedom 
in a calculation of the Sivers function. 
In this context we study the  effects of color by considering the FSIs
with Abelian and non-Abelian gluon
interactions. We confirm the large $N_c$ QCD scaling behavior
of  Sivers functions and further
uncover the deviations for finite $N_c$.
Within this  framework of FSIs we  perform
a quantitative check of approximate relations between T-odd
TMDs and GPD which goes beyond the discussion of overall signs. 
\end{abstract}

\maketitle


Over the past two decades the transverse partonic structure of hadrons
has been the subject of a great deal of theoretical and experimental
study. Central to these investigations are the  observations
of large transverse single spin asymmetries (TSSAs) and azimuthal asymmetries 
in  hadronic reactions--
 from inclusive hadron production~\cite{Adams:2003fx,Adler:2005in,
Abelev:2008qb} to Drell-Yan Scattering~\cite{Guanziroli:1987rp,Conway:1989fs}, 
and in semi-inclusive deep inelastic lepton-nucleon scattering experiments~\cite{Air:2009ti,Alex:2008dn}.  
Two explanations to account for TSSAs in QCD have emerged which
are based on the twist-three~\cite{Efremov:1983eb,Efremov:1984ip,Qiu:1991pp} 
and twist-two~\cite{Sivers:1989cc,Collins:1992kk,Boer:1997nt} approaches.   
We focus on the twist--two approach
 in the factorized picture of semi-inclusive deep inelastic lepton-hadron
scattering (SIDIS)~\cite{Boer:1997nt,Ji:2004wu}
at small transverse momenta of the produced hadron, $P_T\sim k_T << \sqrt{Q^2}$, where  $\sqrt{Q^2}$ is the hard scale. In this kinematic regime the Sivers effect describes a twist-two   transverse target
spin asymmetry
  through the {}``naive'' time reversal odd (T-odd) 
structure, $\Delta f(x,\vec{k}_{T})\sim S_{T}\cdot(P\times\vec{k}_{T})f_{1T}^{\perp}(x,k_{T}^{2})$~\cite{Sivers:1989cc}.  $k_{T}$  is the quark intrinsic transverse momentum and $P$ is
the momentum of the target.   The Sivers asymmetry has been the focus of much theoretical work on QCD factorization theorems. 
Among the most interesting  results 
is that the Sivers function is not universal. It is predicted
that there is a  relative sign between the Sivers function 
from consideration of the gauge link dependence 
going from SIDIS to Drell-Yan 
scattering~\cite{Brodsky:2002rv,Collins:2002kn}. 
The non-universality is further reflected 
in azimuthal asymmetries where one encounters the transverse moments of
the quark correlator $\Phi$~\cite{Boer:2003cm}. Here $\Phi_{\partial}^{i}(x)=\int d^{2}k_{T}\, k_{T}^{i}\,\Phi(x,\vec{k}_{T})$, 
where the non-trivial link dependence remains after integration over
$k_T$.   The correlator decomposes as
$\Phi_{\partial}^{\alpha}(x)=\tilde{\Phi}_{\partial}^{\alpha}(x)+C_{G}\,\pi\Phi_{G}^{\alpha}(x,x)$  with calculable process-dependent gluonic pole factors $C_{G}$~\cite{Bomhof:2006ra} 
(\eg  ${C_{G}}^{(SIDIS)}=-{C_{G}}^{(DY)}$ whereas ${C_{G}}^{(SIDIS)}=1$).   
$\tilde{\Phi}_{\partial}$ contains
the T-even operator combination, while $\Phi_{G}$ contains the T-odd
operator combination. The latter is precisely the soft limit $x_{1}\rightarrow0$
of a quark-gluon correlator $\Phi_{G}(x,x_{1})$. 
It was shown in Ref.~\cite{Boer:2003cm} that the $k_{T}$-weighted 
Sivers function can be written in terms of the gluonic pole matrix 
element which we express 
\be
\langle k_T^i\rangle (x)\hspace{-.05cm}=
\hspace{-.05cm}\int\frac{dz^{-}}{2\pi}\hspace{-.01cm}\mathrm{e}^{ixP^{+}z^{-}}
\hspace{-.05cm}\langle P, S_T|\bar{q}(-z^-\, n/2)\gamma^{+}[-z^-\,n/2;z^-\,n/2]\hat{I}^{i}(z^-\,n/2) q(z^-\,n/2)|P, S_T\rangle\label{eq:RelMI},
\ee
where $\langle k_T^i\rangle (x)\equiv 2M\epsilon_{T}^{ji}S_{T}^{j}f_{1T}^{\perp(1)}(x)$.  The light-like vector $n=(n^-=1,n^+=0,n_\perp=0)$ 
represents a specific direction on the light-cone,  
and $[x\,;\, y]$ denotes a gauge link operator 
connecting the two locations $x$ and $y$. The operator 
$\hat{I}$ originates from the time-reversal behavior of the FSI/ISI implemented by the gauge link operator in (\ref{eq:RelMI}) 
 and is expressed in terms of the gluonic field strength tensor
\begin{equation}
2\hat{I}^{i}(z^-\,n/2)=\int dy^{-}\,[z^-\,n/2,y^{-}n] gF^{+i}(y^{-}n)\,[y^{-}n,z^-\,n/2].
\end{equation}
We exploit the model dependent relation of transverse distortion 
in momentum space with transverse distortion in impact parameter 
space~\cite{Meissner:2007rx} 
by inserting a complete  set of momentum states and demand 
that the operator $\hat{I}$ is diagonal in 
momentum eigenstates, that is
$
\sum_{\lambda^\prime_{\mathbf{P}}}\sum_{\lambda_{\mathbf{P}}}\dots
\langle 
\lambda^\prime_{\mathbf{P}}|\hat{I}^i |\lambda_{\mathbf{P}}\rangle\delta^{(4)}(
\lambda^\prime_{\mathbf{P}}-\lambda_{\mathbf{P}})
$.
\begin{figure}
\includegraphics[width=.45\textwidth]{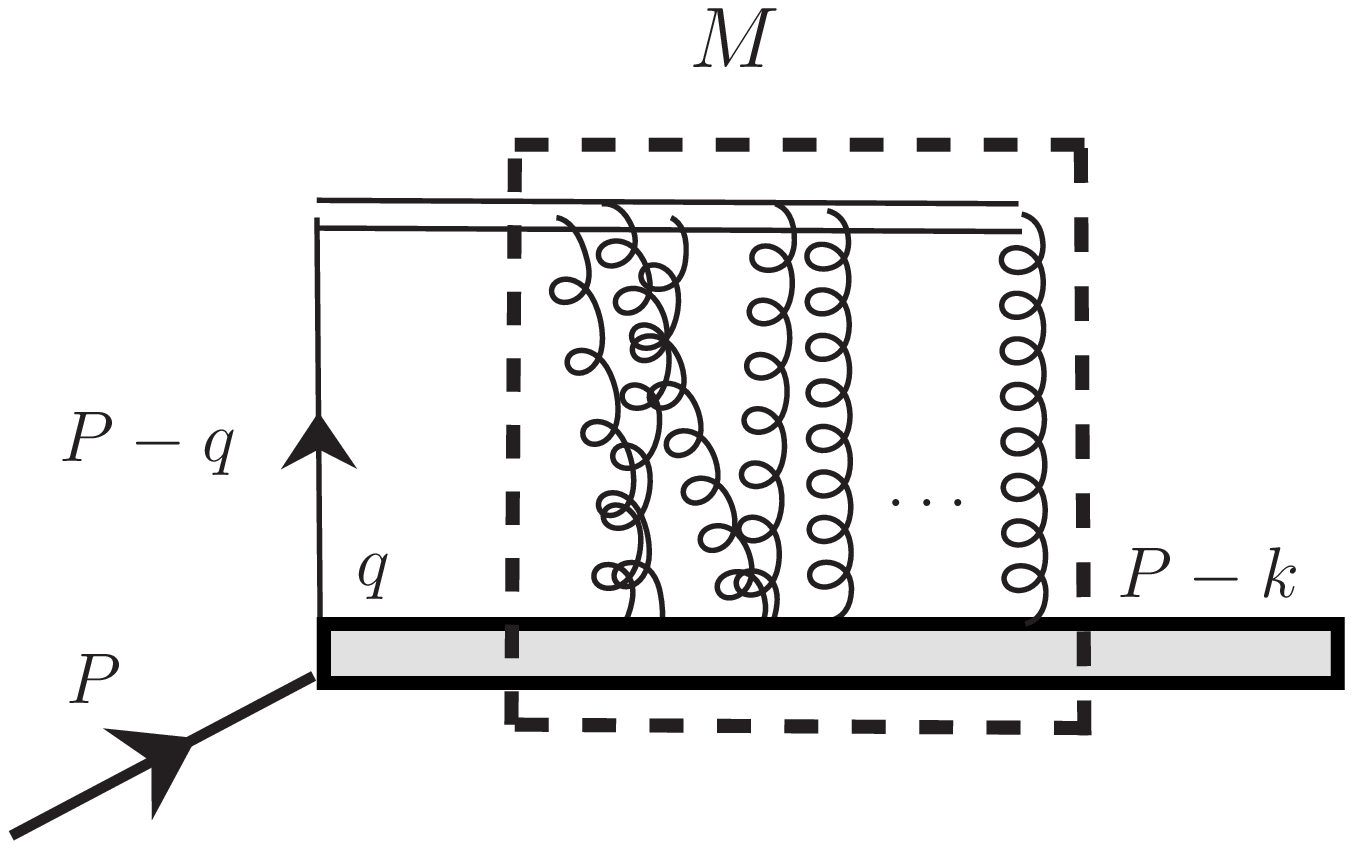}
\hspace{0.20cm}
\includegraphics[width=.55\textwidth]{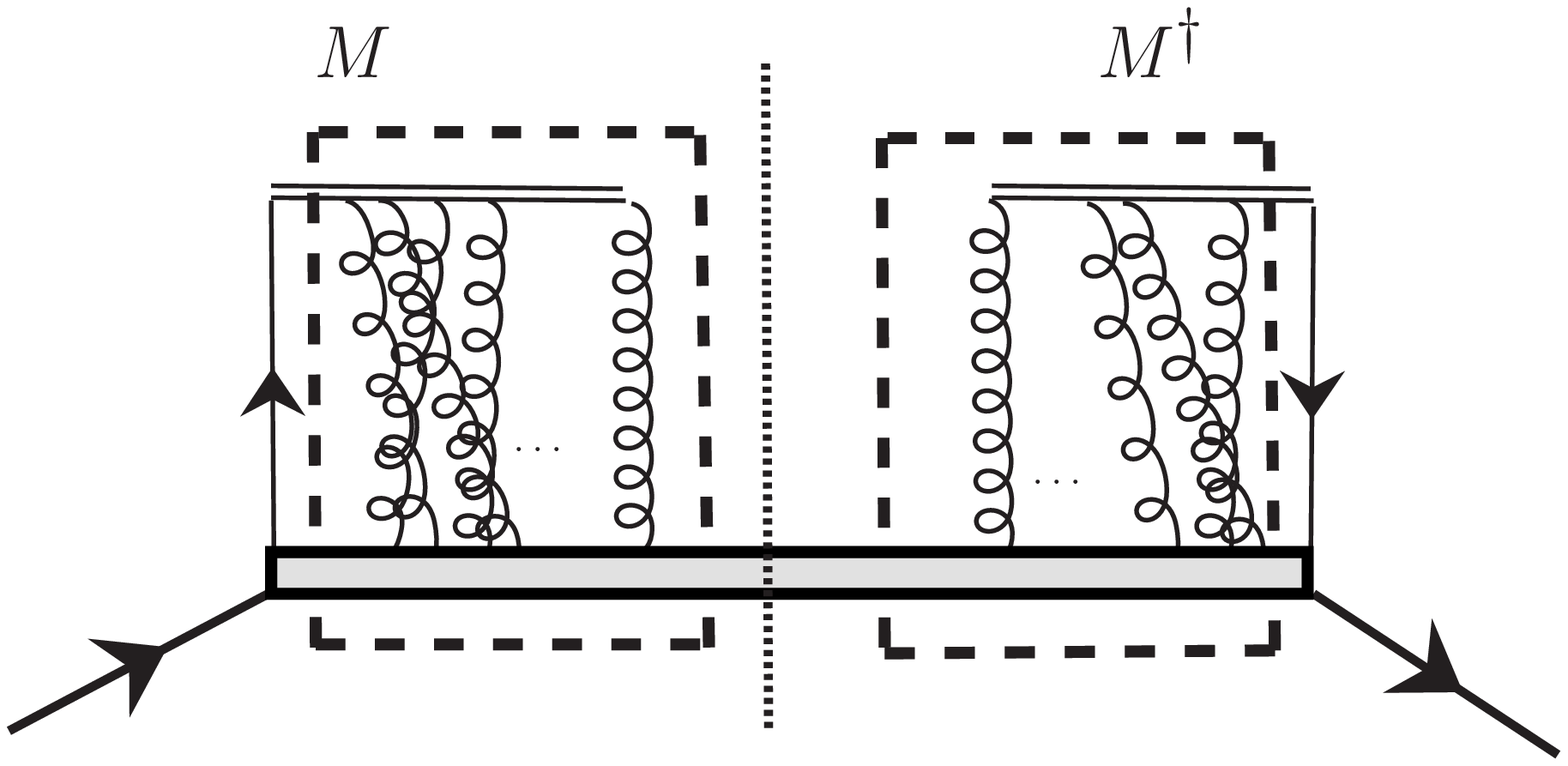}
\caption{Left: The matrix element $W=\langle P-k| \left[\infty\,\ n;0\right] q(0)|P\rangle $ dressed with the FSIs.  The FSIs are described 
by a non-perturbative scattering amplitude $M$ that is calculated in
a generalized ladder approximation~\cite{Gamberg:2009uk}.
Right:  The quark-quark correlator with FSIs. 
}
\label{fig1}
\end{figure}
This leads us to model the target remnant of the 
quark-quark correlation function in terms of sum of spectators states
$\{|\lambda_{\mathbf{P}}\rangle\}$ and to define the lensing function 
$I(\mathbf{P})\equiv
\sum_{\lambda^\prime_{\mathbf{P}}}\langle\lambda^\prime_{\mathbf{P}}|\hat{I} |\lambda_{\mathbf{P}}\rangle\delta^{(4)}(
\lambda^\prime_{\mathbf{P}}-\lambda_{\mathbf{P}})$.  In this picture is it clear why  spectator models
can describe   the gluonic pole matrix element as the factorization of
 FSIs convoluted with transverse impact parameter distortion (via
impact parameter GPDs). While  one  cannot  achieve a general factorization
between TMDs and impact
parameter GPDs; that is for 
the whole sum, the phenomenological relation may exist
for each Fock space state separately. 
  In this approximation  
the Dirac structure  is carried by the distortion and the color dependence
is determined by the chromodynamic lensing function~\cite{Burkardt:2003uw}.
This is most clearly displayed by
transforming Eq.~(\ref{eq:RelMI}) into a mixed coordinate-momentum representation where the  gluonic pole matrix element is expressed as
\begin{equation}
\langle k_T^i\rangle (x)
=\int d^{2}b_{T}\int\frac{dz^{-}}{(2\pi)}\,\mathrm{e}^{ixP^{+}z^{-}}\langle P^{+},\vec{0}_{T};\, S_{T}|\,\bar{q}(z_{1})\gamma^{+}[z_{1}\,;\, z_{2}]\, \hat{I}^{i}(z_{2})\, q(z_{2})\,|P^{+},\vec{0}_{T};\, S_{T}\rangle.\label{eq:GPME}
\end{equation}
The impact parameter $b_{T}$ sits in the arguments of
the quark fields, $z_{1/2}=(\mp\frac{z^{-}}{2\sqrt{2}},\vec{b}_{T},\pm\frac{z^{-}}{2\sqrt{2}})$ and the lensing operator is  $2\hat{I}^{i}(z_2)=\int dy^{-}\,[z_2\,;\, y]\, gF^{+i}(y)\,[y\,;\,z_2]$ where, $y^\mu=y^- n^\mu+b_T^\mu$. For each Fock space state we have the 
phenomenological relation,
\begin{equation}
\langle k_T^i\rangle (x)\simeq2\int d^{2}b_{T}\,{I}^{i}(x,\vec{b}_{T})\frac{\epsilon_{T}^{ij}b_{T}^{i}S_{T}^{j}}{M}\,\frac{\partial}{\partial\vec{b}_{T}^{2}}\mathcal{E}(x,\vec{b}_{T}^{2}).\label{eq:Relation}
\end{equation}
 We utilize  this picture 
to calculate  the gluonic pole matrix element 
using a  soft 
approximation for the lensing function~\cite{Gamberg:2009uk,eikref} which we then convolute
with the parameterizations of impact parameter GPD 
$\mathcal{E}$~\cite{Guidal:2004nd,Diehl:2004cx}.

First we model the target remnant
 in terms of the sum of spectators the quark correlation function 
\begin{equation}
\Phi_{ij}(x,\vec{k}_{T})  =  
\frac{1}{2(2\pi)^{3}(1-x)P^{+}}\sum_{\sigma,\beta}\hspace{-0.45cm}\int
\bar{W}^{\beta \gamma}_j(P,k;\sigma) 
{W}^{\gamma\beta}_i(P,k;\sigma) 
\label{eq:PhiSpectator},
\end{equation}
which is expressed in terms of the matrix element  
$ W_i^{\alpha, \delta}(P,k;\sigma)=\langle P-k,\sigma,\delta|\,[\infty n\,;\,0]^{\alpha \beta}\, q_{i}^{\beta}(0)\,|P\rangle$ where
 $\sigma$ and $\delta$ represent the helicity and color of the intermediate 
states (see Fig.~\ref{fig1}).  
 The FSIs\--- generated by the gauge link 
\--- are described by a non-perturbative amputated scattering amplitude
$({M})_{\gamma \delta}^{\alpha\beta}$ with $\beta,\,\alpha$
($\gamma,\,\delta$) color indices of incoming eikonal quark
and out going spectator remnant. In momentum space 
$W$ is given in terms of $M$
\begin{equation}
\Delta W_{i}^{\alpha\beta}(P,k,S) =  
 \int\frac{d^{4}q}{(2\pi)^{4}}\, ig_{N}\left((P-q)^{2}\right)\frac{\left[(\slash P-\slash q+m_{q})u(P,S)\right]_{i}\left(M\right)_{\delta\beta}^{\alpha\delta}(q,P-k)}{\left[n\cdot(P-k-q)+i0\right]\left[(P-q)^{2}-m_{q}^{2}+i0\right]\left[q^{2}-m_{s}^{2}+i0\right]}\, ,
\label{eq:StartW}
\end{equation}where $i(n\cdot(P-k-q)+i0)^{-1}$ represents the eikonal 
propagator and  $\Delta W=W-W^0$, where $W^0$ 
 denotes the contribution
without final-state interactions. Tracing Eq.~\ref{eq:PhiSpectator} 
with $\gamma^+$ and  weighting, and integrating 
with respect to $k_T$ yields the first moment of the Sivers function~\cite{eikref}
\begin{equation}
\epsilon_{T}^{ij}S_{T}^{j}f_{1T}^{\perp,(1)\,\mathrm{sc}}(x)=-\frac{1}{2(1-x)M^{2}}\int\frac{d^{2}k}{(2\pi)^{2}}\,\epsilon_{T}^{rl}k^{r}S_{T}^{l}I^{i}(x,\vec{k})E(x,0,-\left(\frac{\vec{k}}{1-x}\right)^{2})
\, , \label{eq:RelationDiquark}\end{equation}
where the lensing function $I^{i}$ can be expressed in terms of
the real and imaginary part of the scattering amplitude ${M}$~\cite{Gamberg:2009uk}. 

We use functional methods to incorporate 
the color degrees of freedom for soft gauge boson coupling to highly
energetic particles on the light-cone in the calculation of  $M$, which is 
 given by the expression  
\begin{eqnarray}
\frac{\left(M^{\mathrm{eik}}\right)_{\delta\beta}^{\alpha\delta}(x,|\vec{q}_{T}+\vec{k}_{T}|)}{2(1-x)P^{+}}  = \int d^{2}z_{T}\,\mathrm{e}^{-i\vec{z}_{T}\cdot(\vec{q}_{T}+\vec{k}_{T})}
 \Bigg[\int d^{N_{c}^{2}-1}\alpha\int\frac{d^{N_{c}^{2}-1}u}{(2\pi)^{N_{c}^{2}-1}}\,\mathrm{e}^{-i\alpha\cdot u}\left(\mathrm{e}^{i\chi(|\vec{z}_{T}|)t\cdot\alpha}\right)_{\alpha\delta}\left(\mathrm{e}^{it\cdot u}\right)_{\delta\beta}-\delta_{\alpha\beta}\Bigg].
\label{eq:eikonalAmplitude}
\end{eqnarray}
In Eq.~(\ref{eq:eikonalAmplitude}) the $N_{c}^{2}-1$  dimensional integrals over
the color parameters  results from  auxiliary fields $\alpha^{a}(s)$
and $u^{a}(s)$ that were introduced in the functional formalism of 
Ref.~\cite{Fried:2000hj} in order to decouple the  gluon
fields from the color matrices.   
The eikonal phase $\chi(|\vec{z}_{T}|)$ in Eq.~(\ref{eq:eikonalAmplitude})
represents the amount of soft gluon exchanges that are summed
up into an exponential form, and is given  in terms of the gluon
propagator
\begin{figure}
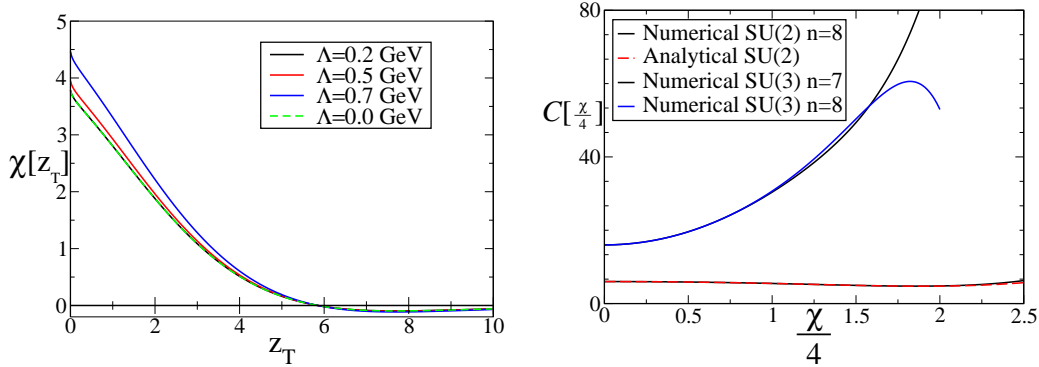

\includegraphics[width=.4\textwidth]{phase.eps}
\hspace{0.3cm}
\includegraphics[width=.4\textwidth]{SU2_3Color.eps}
\caption{Left:The eikonal phase $\chi^{DS}(|\vec{z}_{T}|)$ vs. $|\vec{z}_{T}|$
with input from Dyson-Schwinger equations at scales $\Lambda_{QCD}=0\,\mathrm{GeV},\,0.2\,\mathrm{GeV},\,0.5\,\mathrm{GeV},\,0.7\,\mathrm{GeV}$~\cite{Gamberg:2009uk}. Right: The function $C[\frac{\chi}{4}]$ of Eq.~(\ref{eq:Cchi}) as a function
of the eikonal amplitude $\frac{\chi}{4}$. 
We compare the numerical result computed by means of Eq.~(\ref{eq:PowerSeries})
up to the order $n=8$ with the analytical result 
for the $SU(2)$ color case. The numerical  and  
analytical result agree up to $\frac{\chi}{4}\sim2$.
For $SU(3)$, we compare the numerical results for the
orders $n\,=7,\,8$. The results are reliable for
 $\frac{\chi}{4}\sim1.5$.}
\label{fig2}
\end{figure}
\begin{equation}
\chi(|\vec{z}_{T}|)=g^{2}\int_{-\infty}^{\infty}d\alpha\int_{-\infty}^{\infty}d\beta\, n^{\mu}\bar{n}^{\nu}\mathcal{D}_{\mu\nu}(z+\alpha n-\beta\bar{n}).\label{eq:PhaseEik}\end{equation}
$\mathcal{D}$ denotes the gluon propagator, and $g$ the strong
coupling. In this form the four-vector $v$ is related to 
the complementary light cone vector $\bar{n}$, 
$v=-((1-x)P^{+}/m_{s})\bar{n}$,
with $n\cdot\bar{n}=1$ and $\bar{n}^{2}=0$. 
We  evaluate the color integrals by deriving a power series representation
for the expression in brackets in Eq.~(\ref{eq:eikonalAmplitude}), the
 color function $f_{\alpha\beta}(\chi)$
\begin{equation}
f_{\alpha\beta}(\chi)=\sum_{n=1}^{\infty}\frac{(i\chi)^{n}}{(n!)^{2}}\sum_{a_{1}=1}^{N_{c}^{2}-1}...\sum_{a_{n}=1}^{N_{c}^{2}-1}\sum_{P_{n}}\left(t^{a_{1}}...t^{a_{n}}t^{a_{P_{n}(1)}}...t^{a_{P_{n}(n)}}\right)_{\alpha\beta},
\label{eq:PowerSeries}
\end{equation}
where $P_{n}$ represents the sum over all permutations of the set $\{1,...,n\}$.
If we had  a direct ladder where gluons were not
allowed to cross we would 
have only factors $(t^{a_{1}}...t^{a_{n}}t^{a_{n}}...t^{a_{1}})_{\alpha\beta}=C_{F}^{n}\delta_{\alpha\beta}$
with $C_{F}={N_{c}^{2}-1}/{2N_{c}}$, and we could work in an
Abelian theory with an effective replacement $\alpha\rightarrow C_{F}\alpha_{s}$
for the fine-structure constant. Since we allow generalized ladders
with crossed gluons we have to sum over all permutations in (\ref{eq:PowerSeries}),
and the simple  replacement is not possible. In a large $N_{c}$ expansion
the crossed gluons diagrams would be suppressed such that the direct
ladder represents the leading order in $1/N_{c}$.  
In an Abelian theory, the generating 
matrices $t$ reduce to identity and since we have $n!$ permutations of the
set $\{1,...,n\}$, we  recover the well-known Abelian result,
$f^{\scriptscriptstyle{U(1)}}(\chi)=\sum_{n=1}^{\infty}{(i\chi)^{n}}/{n!}=\mathrm{e}^{i\chi}-1$.  For $N_{c}=2$, $t^{a}=\sigma^{a}/2$
and we can calculate the integral analytically.  We obtain,
$f_{\alpha\beta}^{\scriptscriptstyle{SU(2)}}\left ({\chi}/{4} \right )=\delta_{\alpha\beta}
\left(\cos {\chi}/{4}\,-\,{\chi}/{4}\sin{\chi}/{4}-1\right)+i\delta_{\alpha\beta}\left(2\sin{\chi}/{4}+{\chi}/{4}\cos{\chi}/{4}\right)$.  
We also calculate numerically the lowest coefficients in the power
series (\ref{eq:PowerSeries}), and they  agree with the coefficients
in an expansion in $\chi$ of the analytical result.
This serves as a check of both numerical and analytical
approaches. For  $N_{c}=3$,  
due to  difficulty of
integrating over the Haar measure we use the power series (\ref{eq:PowerSeries}) to obtain  the approximative color function which is valid
when $a=\chi/4$ is small,
\begin{eqnarray}
\Re[f_{\alpha\beta}^{SU(3)}](a)=
 \delta_{\alpha\beta}(-c_{2}a^{2}+c_{4}a^{4}-c_{6}a^{6}-c_{8}a^{8}+...),
\quad
\Im[f_{\alpha\beta}^{SU(3)}](a) = \delta_{\alpha\beta}(c_{1}a-c_{3}a^{3}+c_{5}a^{5}-c_{7}a^{7}+...),\label{eq:CFSU3Im}
\end{eqnarray}
with the numerical values 
$c_{1}=5.333$, $c_{2}=6.222$, $c_{3}=3.951$,
$c_{4}=1.934$, $c_{5}=0.680$, $c_{6}=0.198$, $c_{7}=0.047$, $c_{8}=0.00967$.  
Transforming to  coordinate space  we  can express the lensing 
function directly in terms of the real and imaginary part of the
color function $f$ which is itself a function of the eikonal
phase $\chi$ Eq.~(\ref{eq:PhaseEik}). This results in 
 a lensing function of the form 
\begin{equation}
  {{I}}^{i}(x,\vec{b}_{T})=
\frac{(1-x)}{2N_{c}} \frac{b_{T}^{i}}{|\vec{b}_{T}|}
\frac{\chi^{\prime}}{4} C \left [\frac{\chi}{4} \right ] \  ,\label{eq:LensColor}\end{equation}
where we define the  function, 
\begin{eqnarray}
 C\left [\frac{\chi}{4} \right ]\equiv 
 \Bigl [\left(\Tr\Im[f]\right)^{\prime}\left (\frac{\chi}{4} \right )+ 
\frac{1}{2}\Tr\left[\left(\Im[f]\right)^{\prime}\left (\frac{\chi}{4} \right )
\left(\Re[f]\right)\left (\frac{\chi}{4} \right )\right]
-\frac{1}{2}\Tr\left[\left(\Im[f]\right)\left (\frac{\chi}{4} \right )
\left(\Re[f]\right)^{\prime}\left (\frac{\chi}{4} \right )\right]\Bigr]
 \ ,\label{eq:Cchi}
\end{eqnarray}
and $\chi^{\prime}$ denotes the first derivative with respect to
$|\vec{z}_{T}|$, and $\left(\Im[f]\right)^{\prime}$ and $\left(\Re[f]\right)^{\prime}$ are the first derivatives of the real and imaginary parts of the color
function $f$. 
\begin{figure}[top]
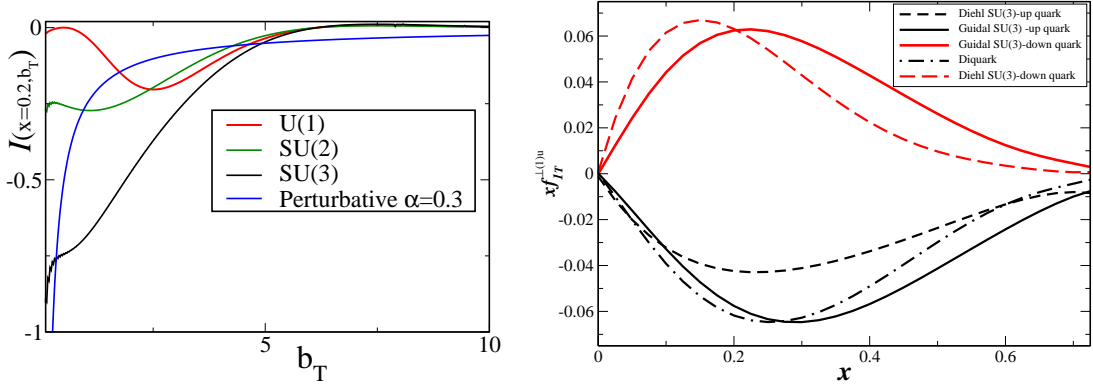

\includegraphics[width=.4\textwidth]{LensingIP.eps}
\hspace{0.25cm}
\includegraphics[width=.45\textwidth]{Sivers_MENU.eps}
\caption{Left:
The lensing function ${I}^{i}(x,\vec{b}_{T})$
from Eq.~(\ref{eq:LensColor}) for U(1), SU(2) and SU(3)
for $x=0.2$ at a scale $\Lambda_{QCD}=0.2\,\mathrm{GeV}$. For comparison
we also plot the perturbative result of Ref.~\cite{Meissner:2008ay}
including the eikonalized antiquark spectator
with an arbitrary value for the coupling,  $\alpha=0.3$. Right: The first
moment of the  Sivers function versus $x$ using various models for the
GPD $E$.}
\label{fig3}
\end{figure}

In Fig.~\ref{fig3}
the function $C[\frac{\chi}{4}]$ is plotted versus $\frac{\chi}{4}$
for various approximations. While the convergence of the power series
seems to be better for $SU(2)$ than in the $SU(3)$ case where the
numerical result calculated with eight coefficients agrees with the
analytical result up to $\frac{\chi}{4}\sim2$, we can trust
 the numerical result computed with eight coefficients up
to $\frac{\chi}{4}\sim1.5$ for $SU(3)$.

In order to numerically estimate the lensing function
and in turn the Sivers function we
utilize the infrared behavior of the gluon
and the running coupling in the non-perturbative regime where we
  infer
that the soft gluon transverse momentum defines the scale at which
the coupling is evaluated. 
These two quantities 
have been extensively studied in the infrared limit in the
 Dyson-Schwinger framework~\cite{Alkofer:2008tt}
and in lattice QCD~\cite{Sternbeck:2008mv}. 
We use calculations of these quantities from  Dyson-Schwinger equations~\cite{Alkofer:2008tt}
where both $\alpha_s$ and $\mathcal{D}^{-1}$ 
are defined in the infrared limit (details can be found in a forthcoming publication).
This determines the eikonal phase and thus the lensing
functions (\ref{eq:LensColor}) for a $U(1)$, $SU(2)$ and $SU(3)$
color function. We plot the results in Fig.~\ref{fig3}
for a color function for $U(1)$, $SU(2)$, $SU(3)$. While we observe
that all lensing functions are attractive and 
fall off at large transverse distances,
they are very different in size at small distances.  

Using the eikonal model for the lensing function together with
different models for the GPD $E$ we present the first moment of the Sivers
function from Eq.~(\ref{eq:Relation}).   
In Fig.~(\ref{fig3})  we compare the diquark model result~\cite{eikref},
to two phenomenological models~\cite{Diehl:2004cx,Guidal:2004nd}. 
We  find that the lensing function grows with $N_c$ which in turn predicts
the growth of the Sivers function. This is consistent with
 the large $N_c$ QCD scaling behavior
of  Sivers functions~\cite{Pobylitsa:2003ty} and further we 
uncover the deviations for finite $N_c$.
 Further, the sign and size of the Sivers function
is consistent with present extractions from data~\cite{Anselmino:2008sga}.
Within this  framework of FSIs we  have performed 
quantitative analysis of approximate relations between T-odd
TMDs and GPD which goes beyond the discussion of overall signs.

\begin{theacknowledgments} L.G. thanks the organizers of MENU for the opportunity to present this work. L.G. acknowledges support from the U.S. Department of 
Energy under contract DE-FG02-07ER41460.
\end{theacknowledgments}

\bibliographystyle{aipproc} 

\bibliography{gamberg_menu_10}


\end{document}